\documentstyle[preprint,eqsecnum,aps]{revtex}
\begin{document}
\title{Reply to ``Comment on `Background thermal contributions in
testing the Unruh effect' ''}
\author{Sandro S.\ Costa and George\ E.A.\ Matsas}
\address{
Instituto de F\'\i sica Te\'orica, Universidade Estadual Paulista\\
R. Pamplona 145, 01405-900 - S\~ao Paulo, S\~ao Paulo, Brazil}
\maketitle


\begin{abstract}
Park et al's recent comment 
that for detectors with large energy gap in
comparison with the temperature of the background
thermal bath, the maximum excitation rate is obtained for some
non-zero detector's velocity  
is {\em correct} but was previously discussed by {\em ourselves} 
in \cite{CM2}, and  does not affect in \cite{CM1} {\em any}
mathematical formula, numerical result, or our
final conclusion that the background thermal bath does not
contribute substantially in the depolarization of electrons at
LEP.

\end{abstract}
\pacs{04.60.+n, 03.70+k}

\narrowtext
Park et al's comment \cite{Petal}
that for detectors with large energy gap $\Delta E$ in
comparison with the temperature $\beta^{-1}$ of the background
thermal bath, one should not state that the faster a detector
moves the less it interacts with the background thermal bath, because
the maximum excitation rate is obtained  for some
detector's velocity $v>0$, 
is true but {\em does not affect in {\em \cite{CM1}} any 
mathematical formula, numerical result, or our final
conclusion that the background thermal bath does not contribute
substantially in the depolarization of electrons at LEP.}
In fact, as calculated in Sec. IV of
\cite{CM1} using a two-level scalar model for the electron, 
the {\em vacuum} contribution  
to the flip probability is three
orders of magnitude larger than the {\em background thermal
bath} contribution, because of the electron acceleration. This 
is consistent with Eq. (2.6) as discussed at the end of Sec.
IV.  For completeness we reproduce Eq. (2.6) of \cite{CM1}
below
$$
\frac{{\cal P}^{\rm exc}}{T^{tot}} = \frac{c^2_0 
\beta^{-1} \sqrt{1-v^2}}{4 \pi v} \ln\left[
         \frac{1-e^{-\beta \Delta E \sqrt{1+v}/\sqrt{1-v}}}
         {1-e^{-\beta \Delta E \sqrt{1-v}/\sqrt{1+v}}}
   \right],
$$ 
where ${\cal P}^{\rm exc}/T^{tot}$ is the excitation rate per proper
time for inertial detectors moving in a background
thermal bath with temperature $\beta^{-1}$, and   
$\Delta E$, $v$ were defined above.
Moreover, we stress that Park et al's
comment above, that for $\Delta E \beta >>1$ 
the maximum excitation rate is obtained  for some
velocity $v>0$ was {\em previously}
stated by {\em ourselves} in Ref. \cite{CM2}
(see caption of Fig. 1), where Eq.
(2.6) was comprehensively discussed in connection with the
classical problem about {\em how temperature transforms under
boosts}. Thus, it seems fair to say that the main (if not sole) 
contribution brought by Park et al's Comment is 
that the $\Delta E$ value in the caption of Fig. 1 of Ref.
\cite{CM1} has a misprint. We have used $\Delta E = 9.7 \times
10^{13} \; s^{-1}$ rather than $\Delta E = 9.7 \times 10^{14} \;
s^{-1}$ to plot this figure {(\em and only this figure)} because of
visual reasons.

In summary, Park et al's comment in \cite{Petal} although
correct was previously discussed by ourselves in
\cite{CM2}, and  does not affect in \cite{CM1} any
mathematical formula, numerical result, or our
final conclusion that the background thermal bath does not
contribute substantially in the depolarization of electrons at
LEP.

\acknowledgments 
S.C. acknowledges full support by Funda\c c\~ao de Amparo \`a
Pesquisa do Estado de S\~ao Paulo, while G.M. was partially
supported by Conselho
Nacional de Desenvolvimento Cient\'{\i}fico e Tecnol\'ogico.

\end{document}